\documentclass[italian,english]{article}
\usepackage[T1]{fontenc}
\usepackage[latin9]{inputenc}
\usepackage{geometry}
\usepackage{color}
\usepackage{amsthm}
\usepackage{amsmath}
\usepackage{amssymb}

\makeatletter

\providecommand{\tabularnewline}{\\}

\newcommand{\lyxaddress}[1]{
\par {\raggedright #1
\vspace{1.4em}
\noindent\par}
}

\makeatother

\usepackage{babel}
\begin{document}

\title{\textbf{Dark Energy and Dark Matter like intrinsic curvature in extended
gravity. Viability through gravitational waves}}

\author{\textbf{Christian Corda }}

\maketitle

\lyxaddress{\begin{center}
Institute for Theoretical Physics and Mathematics Einstein-Galilei,
Via Santa Gonda, 14 - 59100 PRATO, Italy
\par\end{center}}

\begin{center}
\textit{E-mail addresses:} \textcolor{blue}{cordac.galilei@gmail.com}
\par\end{center}
\begin{abstract}
Towards the goal to quantize gravity, in this short review we discuss
an intermediate step which consists in extending the picture of standard
General Relativity by considering \emph{Extended Theories of Gravity}.
In this tapestry, the equations to quantize are not the standard Einstein
field equations of General Relativity, but the\emph{ extended} Einstein
field equations of such Extended Theories\emph{.} The traditional
relation between mass-energy and space-time curvature, which founds
standard General Relativity, results modified in this new picture
and, at least at the linearized approximation, variations of the space-time
curvature generate the mass-energy. 

Various problems of the \emph{Dark Universe}, like \emph{Dark Energy},
\emph{Dark Matter} and \emph{Pioneer anomaly}, can be, in principle,
solved through this approach, while a definitive endorsement for Extended
Theories of Gravity could arrive from the realization of a consistent
\emph{gravitational wave astronomy}. 

We also discuss the quantization of both mass-energy and space-time
curvature in the early Universe by using the process of \emph{amplification
of vacuum fluctuations} which is connected with the primordial production
of relic gravitational waves. A future detection of such relic gravitational
waves will be an ultimate endorsement for the \emph{quantum} rather
than \emph{classical }feature of the gravitational interaction.
\end{abstract}
Although Einstein's General Relativity {[}1{]} achieved great success
(see for example the opinion of Landau who says that General Relativity
is, together with Quantum Field Theory, the best scientific theory
of all {[}2{]}) and withstood many experimental tests, it also displayed
many shortcomings and flaws which today make theoreticians question
whether it is the definitive theory of gravity {[}3{]}. As distinct
from other field theories, like the Electromagnetic Theory, General
Relativity is very difficult to quantize. This fact rules out the
possibility of treating gravitation like other quantum theories and
precludes the unification of gravity with other interactions. At the
present time, it is not possible to realize a consistent \emph{Quantum
Gravity Theory} which leads to the unification of gravitation with
the other forces. 

From a historical point of view, Einstein believed that, in the path
to unification of theories, Quantum Mechanics had to be subjected
to a more general deterministic theory, which he called \emph{Generalized
Theory of Gravitation}, but he did not obtain the final equations
of such a theory (see for example the biography of Einstein which
has been written by Pais {[}4{]}). At present, this point of view
is partially retrieved by some theorists, starting from the Nobel
Laureate G. 't Hooft {[}5{]}. 

During the last 30 years, a strong, critical discussion about both
General Relativity and Quantum Mechanics has been undertaken by theoreticians
in the Scientific Community {[}6, 7, 8, 9{]}. The first motivation
for this historical discussion arises from the fact that one of the
most important goals of Modern Physics is to obtain an \emph{Unified
Theory} which could, in principle, show the fundamental interactions
as different forms of the same \emph{symmetry}. Considering this point
of view, today one observes and tests the results of one or more breaks
of symmetry. In this way, it is possible to say that we live in an
\emph{unsymmetrical} world {[}7, 9{]}. In the last 60 years, the dominant
idea has been that a fundamental description of physical interactions
arises from Quantum Field Theory {[}7, 9{]}. In this approach, different
states of a physical system are represented by vectors in a Hilbert
space defined in a space-time, while physical fields are represented
by operators (i.e. linear transformations) on such a Hilbert space.
The greatest problem is that this quantum mechanical framework is
not consistent with gravitation, because this particular field, i.e.
the metric $g_{mn}$, describes both the dynamical aspects of gravity
and the space-time background {[}6, 7, 9{]}. In other words, one says
that the quantization of dynamical degrees of freedom of the gravitational
field is meant to give a quantum-mechanical description of the space-time.
This is an unequalled problem in the context of quantum field theories,
because the other theories are founded on a fixed space-time background,
which is treated like a classical continuum {[}7, 9{]}. Thus, at the
present time, an absolute Quantum Gravity Theory, which implies a
total unification of various interactions has not been obtained {[}6,
7, 9{]}. In addition, General Relativity assumes a classical description
of the matter which is totally inappropriate at subatomic scales,
which are the scales of the early Universe {[}6, 7, 9{]}.

By considering a geometric unification approach, one assumes that
geometry (for example the Ricci curvature scalar $R$) interacts with
material quantum fields generating back-reactions which modify the
gravitational action adding interaction terms (examples are high-order
terms in the Ricci scalar and/or in the Ricci tensor and non minimal
coupling between matter and gravity, see {[}6, 7, 9{]} and references
within). This approach enables the possibility to extend the framework
of General Relativity {[}6, 7, 9{]} by modifying the Lagrangian, with
respect to the standard Einstein-Hilbert gravitational Lagrangian
{[}2{]}, through the addition of high-order terms in the curvature
invariants (terms like $R^{2}$, $R^{ab}R_{ab}$, $R^{abcd}R_{abcd}$,
$R\Box R$, $R\Box^{k}R$, in the sense of the so-called $f(R)$ \emph{Theories}
{[}9{]}) and/or terms with scalar fields non-minimally coupled to
geometry (terms like $\phi^{2}R$ in the sense of the so-called \emph{Scalar-Tensor
Theories} {[}3, 7, 9{]}, i.e. generalizations of the Jordan-Fierz-Brans-Dicke
Theory {[}10, 11, 12{]}). In general, terms like those are present
in all the approaches to the problem of unification between gravity
and other interactions {[}3, 7, 9{]}. Additionally, from a cosmological
point of view, such modifications of General Relativity generate \emph{Inflationary
frameworks} which are very important as they solve many problems of
the \emph{Standard Universe Model} {[}9, 13{]}.

Thus, in the framework of the Quantum Gravity problem, a possible
extension of General Relativity is important as it represents the
connection between the, at the present time unknown, correct and definitive
Quantum Gravity Theory and Einstein's General Relativity. 

The necessity to produce a correct Quantum Gravity Theory came into
existence at the end of 50's of last century, when scientists tried
to analyse the four interactions at a fundamental level in the sense
of Quantum Field Theory {[}6, 7, 9{]}. The starting point was to follow
the same type of analysis performed considering the other interactions:
for example, the Electromagnetic Theory was quantized following both
of the canonical and covariant approaches. In the first case, one
considers magnetic and electric fields which satisfy the \emph{Uncertainty
Principle} and quantum states which are functions of invariant gauges
generated by potential vectors on 3-surfaces {[}7, 9{]}. Instead,
in the covariant approach, one isolates and quantizes the two degrees
of freedom of the Maxwellian field, without the 3+1 metric decomposition,
and the quantum states are given by elements of the Fock space of
photons {[}7, 9{]}. 

The two cited approaches are equivalent in the case of Electromagnetic
Theory, but, when scientists tried to apply the same analysis to gravitation,
they obtained deep differences {[}6, 7, 9{]}. The biggest difficulty
is the fact that General Relativity cannot be formulated like a Quantum
Field theory on Minkowskian space-time, as in General Relativity a
geometry a priori is not present in the space-time background. In
fact, the space-time is the final product of the evolution, i.e. the
dynamic variable {[}7, 9{]}. Then, if ones wants to introduce fundamental
notions like causality, time and evolution of the system, one has
to solve the field equations obtaining a particular space-time as
solution. Let us consider the classical example of a black hole {[}2,
7, 9{]}. To understand if particular boundary constraints generate
a black hole, one has to solve the Einstein field equations. After
this, by using the causal structure induced by the solution, one has
to study the asymptotic future metric and connect it to the past initial
data. It is very difficult to discuss the problem from a quantum point
of view. The Uncertainty Principle prevents particles having definite
trajectories even in non-relativistic Quantum Mechanics; the time
evolution gives only an amplitude probability rather than a precise
trajectory {[}7, 9{]}. In the same way, in Quantum Gravity, the evolution
of the initial state cannot give a specific space-time. Then, it is
not possible to introduce fundamental concepts like causality, time
and matrix elements. 

The two cited approaches, i.e. the covariant and canonical approaches,
give different solutions to these problems. Substantially, the Quantum
Gravity problem is represented exactly by this inconsistency {[}7,
9{]}. 

In the general context of cosmological evidence, there are also other
considerations which suggest an extension of General Relativity {[}3,
9{]}. As a matter of fact, the accelerated expansion of the Universe,
which is observed today, implies that cosmological dynamics is dominated
by the so called \emph{Dark Energy}, which gives a large negative
pressure. This is the standard picture, in which this new ingredient
should be some form of un-clustered, non-zero vacuum energy which,
together with the clustered \emph{Dark Matter}, drives the global
dynamics. This is the so called \textquotedblleft{}\emph{concordance
mode}l\textquotedblright{} ($\Lambda$CDM) which gives, in agreement
with the \emph{Cosmic Microwave Background Radiation}, \emph{Large
Scale Structure} and \emph{Supernovae Ia} data, a good picture of
the observed Universe today, but presents several shortcomings such
as the well known \textquotedblleft{}\emph{coincidence}\textquotedblright{}
and \textquotedblleft{}\emph{Cosmological Constant}\textquotedblright{}
problems {[}9, 14{]}. 

An alternative approach is seeing if the observed cosmic dynamics
can be achieved through an extension of General Relativity {[}3, 8,
9. In this different context, it is not required to find candidates
for Dark Energy and Dark Matter, that, till now, have not been found;
only the \textquotedblleft{}\emph{observed}\textquotedblright{} ingredients,
which are curvature and baryon matter, have to be taken into account
{[}9{]}. Then, Dark Energy and Dark Matter have to be considered like
pure effects of the presence of an intrinsic curvature in the Universe.
Considering this point of view, one can think that gravity is different
at various scales because of the existence of the intrinsic curvature,
which changes at different scales, and there is room for alternative
theories. 

Note that we are not claiming that General Relativity is wrong. It
is well known that, even in the context of Extended Theories of Gravity,
General Relativity \textit{remains the most important part of the
structure }{[}3, 8, 9{]}. We are only trying to understand if weak
modifies on such a structure, i.e. trying to extend it, could be needed
to solve some theoretical and current observational problems {[}9{]}.
In this picture, we also recall that even Einstein tried to modify
the framework of General Relativity by adding the \textquotedblleft{}\emph{Cosmological
Constant}\textquotedblright{} {[}15{]}. In any case, Solar System
tests show that modifications of General Relativity in the sense of
Extended Theories of Gravity have to be very weak {[}3{]}. In the
geometric approach of this short review such an important point requests
that the scales of the spatial hypersurfaces of the intrinsic curvature
have to be much larger than the dimensions of the Solar System. However,
from the point of view of the goals of this short review, we emphasize
that the various constraints on Extended Theories of Gravity \emph{do
not ban} their viability {[}3, 7, 9{]}. In fact, on cosmological scales
observations indicate that spatial hypersurfaces (orthogonal to cosmic
time) of space-time are flat or almost flat, but space-time could
be globally curved {[}29, 30{]}.

In principle, the most popular Dark Energy and Dark Matter models
can be achieved in the framework of Extended Theories of Gravity {[}7,
8, 9, 16, 17{]}. 

The idea which founds standard General Relativity is that the geometry
of space-time is determined by the distribution of mass-energy, i.e.
following {[}18{]}, ``\emph{Matter tells space how to curve. Space
tells matter how to move}''. The mathematical description is given
by Einstein field equations that, by assuming $G=1$, $c=1$ and $\hbar=1$
(\emph{natural units}), read {[}18{]}

\begin{equation}
G_{ik}=8\pi T_{ik}^{(m)},\label{eq: Einstein}
\end{equation}

where the Einstein tensor $G_{ik}$ represents the geometry of space-time
and the stress-energy tensor $T_{ik}^{(m)}$ represents the distribution
of mass-energy. The framework of Extended Theories of Gravity modifies
Einstein field equations {[}8, 9, 16, 17, 20, 23{]}

\begin{equation}
G_{ik}=8\pi T_{ik}^{(total)}=8\pi(T_{ik}^{(m)}+T_{ik}^{(c)}),\label{eq: Einstein modified}
\end{equation}

where $T_{ik}^{(c)}$ represents a new stress-energy tensor due to
an intrinsic curvature. It depends on the analytic expression, i.e.
on the Lagrangian, of the particular Extended Theory under consideration.
Then, the ultimate theory of gravity should give the correct mathematical
description of the intrinsic curvature through the correct $T_{ik}^{(c)}$. 

Thus, the relation between mass-energy and geometry of space-time
is changed and the new interpretation reads \emph{``the geometry
of space-time is determined by the distribution of mass-energy plus
the intrinsic curvature}''. The modified Einstein field equations
(\ref{eq: Einstein modified}) are obtained with the same standard
variational analysis in {[}2{]}, the difference is that now the Einstein-Hilbert
action of standard General Relativity, which is linear in terms of
the Ricci scalar $R$, is generalized by adding high-order terms in
the curvature invariants and/or terms with scalar fields non-minimally
coupled to geometry, see {[}8, 9, 16, 17, 20, 23{]} for details. The
presence of such terms generate the \emph{curvature stress-energy
tensor} $T_{ik}^{(c)}$ in Eqs. (\ref{eq: Einstein modified}). Clearly,
if an extension of General Relativity is needed, the correct way to
obtain a Quantum Gravity Theory is to quantize the \emph{extended}
Einstein field equations (\ref{eq: Einstein modified}) rather than
the \emph{standard} Einstein field equations (\ref{eq: Einstein})
of General Relativity. 

An important point is the following. Starting from the modified Einstein
field equations (\ref{eq: Einstein modified}), the linearized theory
in vacuum (i.e. with $T_{ik}^{(m)}=0$ in Eqs. (\ref{eq: Einstein modified}))
that has been developed in Refs. {[}8, 16{]} and from {[}19{]} to
{[}24{]},  gives the Klein-Gordon equation

\begin{equation}
\square h_{c}=m^{2}h_{c},\label{eq: mass production}
\end{equation}

where the quantity $h_{c}$ represents the normalized variation of
a function of the Ricci scalar $R$ in the case of $f(R)$ Theories
or of a function of scalar fields in Scalar-Tensor Theories {[}8,
16{]}, {[}19{]} - {[}24{]}. In other words, $h_{c}$ represents a
function of the variation of the intrinsic curvature (the subscript
$c$ in $h_{c}$ means \emph{curvature}). Eq. (\ref{eq: mass production})
defines the mass-energy like

\begin{equation}
m\equiv\sqrt{\frac{\square h_{c}}{h_{c}}}.\label{eq: massa}
\end{equation}

Thus, as the mass-energy is generated by variations of the Ricci curvature
scalar and/or of scalar fields, we can say that it is generated by
variations of the space-time curvature\textbf{ }{[}8, 16{]}, {[}19{]}
- {[}24{]}. The solution of Eq. (\ref{eq: mass production}) is {[}8,
16{]}, {[}19{]} - {[}24{]}

\begin{equation}
h_{c}=a(\overrightarrow{p})\exp(iq^{a}x_{a})+c.c.,\label{eq: wave-packet}
\end{equation}

where $q^{a}\equiv(\omega_{c},\overrightarrow{p}),\mbox{ }\omega_{c}=\sqrt{m^{2}+p^{2}}$
and the solution (\ref{eq: wave-packet}) is discussed like a wave-packet
(the dispersion law for the modes of $h_{c}$ is not linear).

The linearized theory is quite important also from a different point
of view. In fact, a definitive endorsement for the real viability
of Extended Theories of Gravity could arrive from the realization
of a consistent \emph{gravitational wave (GW) astronomy} {[}3{]}.
Let us discuss this point in detail. 

The scientific community hopes in a first direct detection of GWs
in next years {[}25{]}. The realization of a GW astronomy, by giving
a significant amount of new information, will be a cornerstone for
a better understanding of gravitational physics. The discovery of
GW emission by the compact binary system PSR1913+16, composed by two
neutron stars, by Hulse and Taylor {[}26{]}, Nobel Prize winners,
has been, for physicists working in this field, the ultimate thrust
allowing to reach the extremely sophisticated technology needed to
investigate in this field of research. 

For the goals of this sort review, the sensitive detectors for GWs,
like bars and interferometers, whose data analysis recently started
{[}25{]}, will be definitive to confirm in an ultimate way the physical
consistency of General Relativity {[}3{]} or, alternatively, to endorse
Extended Theories of Gravity. In fact, in the context of gravitational
theories, some differences between General Relativity and Extended
Theories, can be pointed out starting from the linearized theory of
gravity {[}3, 8, 20, 23, 35, 37{]}. 

In 1957, F. A. E. Pirani, who was a member of the Bondi's research
group, proposed the geodesic deviation equation as a tool for designing
a practical GW detector {[}27{]}. Pirani showed that if a GW propagates
in a spatial region where two test masses are present, the effect
is to drive the masses to have oscillations. 

In {[}3, 8, 20, 23, 28, 35, 37{]} it has been shown that in the case
of Extended Theories of Gravity GWs generate different oscillations
of test masses with respect to GWs in standard General Relativity.
Thus, an accurate analysis of such a motion can be used in order to
discriminate among various theories. 

In general, GWs manifest them-self by exerting tidal forces on the
test-masses {[}18{]} which are the mirror and the beam-splitter in
the case of an interferometer {[}25{]}. Putting the beam-splitter
in the origin of the coordinate system, the components of the separation
vector between the two test masses are the mirror's coordinates. At
first order in $h_{+}$, which is the weak perturbation due to the
$+$ polarization {[}18{]}, the displacements of the mirror due by
this $+$ polarization for a GW propagating in the $z$ direction
and in the locally inertial coordinate system are given by {[}18{]}

\begin{equation}
\begin{array}{c}
\delta x_{M}(t)=\frac{1}{2}x_{M0}h_{+}(t)\\
\\
\delta y_{M}(t)=-\frac{1}{2}y_{M0}h_{+}(t),
\end{array}\label{eq: spostamenti GR}
\end{equation}

where $x_{M0}$ and $y_{M0}$ are the initial (unperturbed) coordinates
of the mirror. The $\times$ polarization $h_{\times}(t)$ generates
an analogous motion for test masses which are rotated of 45-degree
with respect the $z$ axis {[}18{]}. $h_{+}(t)$ and $h_{\times}(t)$
are the \emph{sole} two polarizations present in standard General
Relativity and are \emph{plane waves }{[}3, 18{]}\emph{.}

In the case of Extended Theories of Gravity,\emph{ a third polarization}
of GWs, which is in general \emph{massive,} is present {[}3, 8, 20,
23, 28, 35, 37{]}. This new polarization is exactly the quantity $h_{c}$
of Eq. (\ref{eq: mass production}) which is generated by perturbations
of the intrinsic curvature, see eq. ( \ref{eq: massa}). 

The most propitious astrophysical sources of this third polarization
of GWs are gravitational collapses of quasi-spherical astrophysical
objects, see {[}28, 36, 37{]} and the appendix of this short review
for details. 

For a sake of completeness we recall that in a particular case of
Scalar-Tensor Theories of Gravity the third mode can have \emph{zero
mass}, i.e. Eq. (\ref{eq: mass production}) becomes {[}3, 23, 28,
35, 37{]}

\begin{equation}
\square h_{c}=0.\label{eq: zero-mass}
\end{equation}
 In that case, perturbations of the intrinsic curvature generate massless
particles (waves) and the solution (\ref{eq: wave-packet}) becomes
an ordinary plan wave propagating at the speed of light {[}3, 23,
28, 35, 37{]}. In any case, we recall that these particles carry energy
away from their astrophysical sources {[}28, 37{]}, thus, even being
massless, such particles have an associated energy. At first order
in $h_{c}$, the displacements of the mirror due to this third massless
polarization in the locally inertial coordinate system are given by
{[}23, 28, 35{]}

\begin{equation}
\begin{array}{c}
\delta x_{M}(t)=\frac{1}{2}x_{M0}h_{c}(t)\\
\\
\delta y_{M}(t)=\frac{1}{2}y_{M0}h_{c}(t).
\end{array}\label{eq: spostamenti massless}
\end{equation}

Thus, the total motion of the mirror due to GWs in this case is (third
massless mode plus \emph{``ordinary}'' modes which are present in
standard General Relativity too, we use the principle of superposition
of modes) {[}23, 28, 35{]} 
\begin{equation}
\begin{array}{c}
\delta x_{M}(t)=\frac{1}{2}x_{M0}h_{+}(t)+\frac{1}{2}x_{M0}h_{c}(t)\\
\\
\delta y_{M}(t)=-\frac{1}{2}y_{M0}h_{+}(t)+\frac{1}{2}y_{M0}h_{c}(t).
\end{array}\label{eq: spostamenti massless totali}
\end{equation}

The case of massive waves is more general as it is present in all
$f(R)$ Theories and in almost all Scalar-Tensor Theories {[}3, 8,
23, 28, 35{]}. In that case, the displacements of the mirror due to
the third massive polarization in the locally inertial coordinate
system are {[}8, 23, 28, 35{]}

\begin{equation}
\begin{array}{c}
\delta x_{M}(t)=\frac{1}{2}x_{M0}h_{c}(t)\\
\\
\delta y_{M}(t)=\frac{1}{2}y_{M0}h_{c}(t)\\
\\
\delta z_{M}(t)=\frac{m^{2}}{2\omega_{c}^{2}}z_{M0}h_{c}(t),
\end{array}\label{eq: spostamenti}
\end{equation}

where $\omega_{c}$ is the frequency of the propagating massive GW.
Thus, the total motion of the mirror due to GWs in the massive case
is (third massive mode plus \emph{``ordinary}'' modes) {[}28, 35{]}

\begin{equation}
\begin{array}{c}
\delta x_{M}(t)=\frac{1}{2}x_{M0}h_{+}(t)+\frac{1}{2}x_{M0}h_{c}(t)\\
\\
\delta y_{M}(t)=-\frac{1}{2}y_{M0}h_{+}(t)\frac{1}{2}y_{M0}h_{c}(t)\\
\\
\delta z_{M}(t)=\frac{m^{2}}{2\omega_{c}^{2}}z_{M0}h_{c}(t).
\end{array}\label{eq: spostamenti totali massive}
\end{equation}

Then, in the case of massive GWs a longitudinal component, which is
due to the presence the small mass $m,$ is present {[}8, 23, 28,
35{]}. 

Thus, if advanced projects on the detection of GWs will improve their
sensitivity allowing to perform a GW astronomy (this is due because
signals from GWs are quite weak {[}3, 25{]}), one will only have to
look which is the motion of the mirror in respect to the beam splitter
of an interferometer in the locally inertial coordinate system in
order to understand if Extended Theories of Gravity are viable. If
such a motion will be governed only by Eqs. (\ref{eq: spostamenti GR})
we will conclude that General Relativity is the ultimate Theory of
Gravity. If the motion of the mirror is governed by Eqs. (\ref{eq: spostamenti massless totali})
(massless case) or by Eqs. (\ref{eq: spostamenti totali massive})
(massive case), we will conclude that we need to extend standard General
Relativity in terms of Extended Theories of Gravity.

Even if such signals will be quite weak, a consistent GW astronomy,
which will use coincidences between various detectors, will permit,
in principle, to determine all the quantities of the above equations
{[}3, 8, 23, 28, 35, 37{]}. 

Let us resume the situation by including a Table with 3 rows and 2
columns. In the first column we include the 3 models to be distinguished
(i.e. which is the correct theory among General Relativity, Extended
Theory of Gravity with production of massless particles, Extended
Theory of Gravity with production of massive particles) and in the
second column we include the corresponding motion of the mirror.

\bigskip{}

\subsection*{Table}

\begin{tabular}{|c|c|}
\hline 
Standard General Relativity & $\begin{array}{c}
\\
\delta x_{M}(t)=\frac{1}{2}x_{M0}h_{+}(t)\\
\\
\delta y_{M}(t)=-\frac{1}{2}y_{M0}h_{+}(t)\\
\\
\end{array}$\tabularnewline
\hline 
\hline 
\begin{tabular}{c}
Extended Theory of Gravity with\tabularnewline
production of massless particles\tabularnewline
\end{tabular} & $\begin{array}{c}
\\
\delta x_{M}(t)=\frac{1}{2}x_{M0}h_{+}(t)+\frac{1}{2}x_{M0}h_{c}(t)\\
\\
\delta y_{M}(t)=-\frac{1}{2}y_{M0}h_{+}(t)+\frac{1}{2}y_{M0}h_{c}(t)\\
\\
\end{array}$\tabularnewline
\hline 
\begin{tabular}{c}
Extended Theory of Gravity with\tabularnewline
production of massive particles\tabularnewline
\end{tabular} & $\begin{array}{c}
\\
\delta x_{M}(t)=\frac{1}{2}x_{M0}h_{+}(t)+\frac{1}{2}x_{M0}h_{c}(t)\\
\\
\delta y_{M}(t)=-\frac{1}{2}y_{M0}h_{+}(t)+\frac{1}{2}y_{M0}h_{c}(t)\\
\\
\delta z_{M}(t)=\frac{m^{2}}{2\omega_{c}^{2}}z_{M0}h_{c}(t)\\
\\
\end{array}$\tabularnewline
\hline 
\end{tabular}

\bigskip{}

Another important point is the following. Let us ask: is it possible
interpreting Eqs. (\ref{eq: mass production}) and (\ref{eq: massa})
in order to realize a (linearized) \emph{quantization} of both mass-energy
and space-time curvature? This important issue has been discussed
in {[}19, 21, 24, 32, 33, 34{]} by using the process of \emph{amplification
of vacuum fluctuations} in the early Universe in the framework of
the production of relic GWs in Extended Theories of Gravity. The analysis
arises from a mixing between basic principles of Classical Theories
of Gravity and of Quantum Field Theory {[}19, 21, 24, 32, 33, 34{]}.
It is very important to stress the unavoidable and fundamental character
of the mechanism which produces these particles. The strong variations
of the gravitational field and of the space-time curvature in the
early Universe amplify the zero-point quantum oscillations and produce
particles like the ones of Eq. (\ref{eq: massa}). Again, even if
in general such GWs are massive, we emphasize that in a particular
case (massless Scalar-Tensor Gravity) they can be massless as Eq.
(\ref{eq: mass production}) reduces to Eq. (\ref{eq: zero-mass})
{[}3, 23, 28, 35, 37{]}. The model derives from the Inflationary scenario
for the early Universe {[}13{]}, which is consistent in a good way
with the WMAP data on the \emph{Cosmic Background Radiation} (in particular
exponential Inflation and spectral index $\approx1$ {[}29, 30{]}).
Inflationary models of the early Universe were analysed in the early
and middles 1980's {[}13{]}. These are cosmological models in which
the Universe undergoes a brief phase of a very rapid expansion in
early times. In this context the expansion could be power-law {[}31{]}
or exponential in time {[}19{]}. Inflationary models provide solutions
to the horizon and flatness problems and contain a mechanism which
creates perturbations in all fields {[}13{]}. Important for our goals
is that this mechanism, when applied to Extended Theories of Gravity,
also provides the production of mass-energy, in terms of GWs like
the ones of Eq. (\ref{eq: massa}), from variations of the intrinsic
curvature. The zero-point amplitude of the wave-packet (particle)
$h_{c}$ is derived through the quantization relations {[}32, 33,
34{]} 
\begin{equation}
[h_{c}(t,x),\pi_{h_{c}}(t,y)]=i\delta^{3}(x-y)\label{eq: commutare}
\end{equation}
 The result is that the number of created particles (quanta) by variations
of the intrinsic curvature and of angular frequency $\omega_{c}$
results (see {[}19, 21, 24{]} for details of the computation)

\begin{equation}
N_{\omega_{c}}=(\frac{H_{inf}H_{0}}{2\omega_{c}^{2}})^{2},\label{eq: numero quanti}
\end{equation}

where $H_{0}$ and $H_{inf}$ are respectively the actual and the
inflationary values of the Hubble expansion rate and $\omega_{c}$
is the angular frequency that would be observed today if these primordial
particles will be detected by GWs detectors {[}19, 21, 24{]}. Thus,
variations of curvature which generate the production of mass-energy
through Eq. (\ref{eq: massa}) \emph{result quantized in the relic
Universe} and such a quantization generates a number of primordial
particles which is given by Eq. (\ref{eq: numero quanti}). 

Then, the effective detection of these relic particles (waves) will
be a definitive endorsement for the quantum rather than classical
feature of the gravitational interaction.

\section*{Conclusion remarks}

In the framework of attempts to quantize gravity we discussed an intermediate
step which consists in extending the picture of standard General Relativity
in the framework of Extended Theories of Gravity. 

The traditional relation between mass-energy and curvature, which
founds standard General Relativity, results modified in the new picture
and, at least at the linearized approximation, variations of an intrinsic
curvature generate the mass-energy. 

In the proposed approach, the equations to quantize are not the standard
Einstein field equations of General Relativity, but the\emph{ extended}
Einstein field equations of Extended Theories of Gravity. 

Various problems of the Dark Universe, can be, in principle, solved
through this approach. In fact, Dark Energy, Dark Matter and the Pioneer
anomaly are interpreted like pure curvature effects. 

A definitive endorsement for the real viability of Extended Theories
of Gravity could arrive from the realization of a consistent GW astronomy. 

At the end of this short review we also discussed the quantization
of both mass-energy and curvature in the early Universe by using the
process of amplification of vacuum fluctuations which is connected
with the primordial production of relic GWs. A future detection of
these relic GWs will be an ultimate endorsement for the quantum rather
than classical feature of the gravitational interaction.

\section*{References}

{[}1{]} A. Einstein,\textit{ }\textit{\emph{Preuss. Akad. Wiss. Berlin,
Sitzber., 778 (1915).}}

{[}2{]} L. Landau and E. Lifsits\textit{, Classical Theory of Fields}
(3rd ed.), London: Pergamon (1971). 

{[}3{]} C. Corda, Int. Journ. Mod. Phys. D, 18, 14, 2275 (2009, Honorable
Mention at Gravity Research Foundation).

{[}4{]} A. Pais, \textit{``Subtle Is the Lord: The Science and the
Life of Albert Einstein''}, Oxford University Press (2005).

{[}5{]} G. 't Hooft, quant-ph/06/04/008 (2006).

{[}6{]}\foreignlanguage{italian}{ H. Nicolai, G. F. R. Ellis, A. Ashtekar
and others, \emph{``Special Issue on quantum gravity}'', Gen. Rel.
Grav. 41, 4, 673 (2009).}

{[}7{]}\foreignlanguage{italian}{ }C. Corda, Ph. D Thesis discussed
at the Pisa University, http://etd.adm.unipi.it/theses/available/etd-09092008-185152/

{[}8{]} C. Corda, J. Cosmol. Astropart. Phys. JCAP04009 (2007).

{[}9{]} S. Nojiri and S. D. Odintsov, Int. J. Geom. Meth. Mod. Phys.
4, 115-146 (2007); S. Capozziello and M. De Laurentis, Phys. Rept.
509, 167 (2011); S. Capozziello, V. Faraoni, Beyond Einstein Gravity:
A Survey Of Gravitational Theories For Cosmology And Astrophysics,
Fundamental Theories of Physics 170, Springer, New York (2011); V.
Faraoni and T. P. Sotiriou Rev. Mod. Phys. 82, 451, (2010); A. A.
Starobinsky, JETP Lett., 30, 682, (1979); A. A. Starobinsky, Phys.
Lett. B 91, 99, (1980); A. A. Starobinsky, Sov. Phys. JEPT Lett. B
34, 438 (1982); S. H. Hendi, Phys. Lett. B 690, 220 (2010); S. H.
Hendi, B. Eslam Panah and S. M. Mousavi, Gen. Relativ. Gravit. 44,
835 (2012); S. Nojiri, S. D. Odintsov, Physics Reports 505, 59, (2011);
A. De Felice, S. Tsujikawa, Living Rev. Rel. 13, 3, (2010); G. J.
Olmo, Int. J. Mod. Phys. D 20, 413, (2011); G. J. Olmo, H. Sanchis-Alepuz,
S. Tripathi, Phys. Rev. D 80, 024013, (2009); G. J. Olmo, H. Sanchis-Alepuz,
Phys. Rev. D 83, 104036, (2011); G. Cognola , E. Elizalde , S. Nojiri,
S.D. Odintsov, S. Zerbini, JCAP 0502, 010, (2005); G. Cognola , E.
Elizalde , S. Nojiri, S.D. Odintsov, S. Zerbini, Phys.Rev. D 73, 084007,
(2006); S. Nojiri and S. D. Odintsov, Phys. Lett. B, 657, 238 (2008);
G. Allemandi, A. Borowiec, M. Francaviglia, S. D. Odintsov , Phys.
Rev. D 72, 063505 (2005); G. Allemandi, A. Borowiec, M. Francaviglia,
Phys. Rev. D 70 (2004) 103503; Will C. M. Theory and Experiments in
Gravitational Physics, Cambridge Univ. Press, Cambridge (1993); S.
Capozziello and M. Francaviglia, Gen. Rel. Grav. 40, 2-3, 357 (2008);
S. Capozziello, M. De Laurentis, M. Francaviglia, S. Mercadante, Foundations
of Physics 39, 1161, (2009); S. Capozziello, E. Elizalde, S. Nojiri,
S. D. Odintsov, Phys. Lett. B 671, 193, (2009); E. Elizalde, S. Nojiri,
and S. D. Odintsov , Phys. Rev. D 70, 043539 (2004); E. Elizalde,
P. J. Silva, Phys. Rev. D 78, 061501 (2008); G. Cognola, E. Elizalde,
S. Nojiri, S. D. Odintsov, L. Sebastiani, S. Zerbini , Phys. Rev.
D 77, 046009 (2008) ; K. Bamba, S. Nojiri and S. D. Odintsov, J. Cosmol.
Astropart. Phys. JCAP10(2008)045; T. Clifton, Phys. Rev. D 77, 024041,
(2008); T. Clifton, J. D. Barrow, Phys. Rev. D 72, 103005, (2005);
T. Clifton, J. D. Barrow, Class. Quantum Grav. 23, 2951, (2006); T.P.
Sotiriou, Gen. Relat. Gravit. 38, 1407, (2006); T.P. Sotiriou, Class.
Quant. Grav. 23, 5117, (2006); T. P. Sotiriou, E. Barausse, Phys.
Rev. D 75, 084007, (2007). 

{[}10{]} P. Jordan, Naturwiss. 26, 417 (1938).

{[}11{]} M. Fierz, Helv. Phys. Acta 29, 128 (1956). 

{[}12{]} C. Brans and R. H. Dicke, Phys. Rev. 124, 925 (1961).

{[}13{]}\foreignlanguage{italian}{ D. H. Lyth and A. R. Liddle, ``\emph{Primordial
Density Perturbation}'', Cambridge University Press (2009).}

{[}14{]} P. J. E. Peebles and B. Ratra, Rev. Mod. Phys. 75, 559 (2003).

{[}15{]} A. Einstein, Sitzungsber. Preuss. Akad. Wiss., 142, 235 (1931).

{[}16{]} C. Corda, Gen. Rel. Grav. 40, 2201 (2008).

{[}17{]} C. Corda, Mod. Phys. Lett. A 23, 109 (2008). 

{[}18{]}\foreignlanguage{italian}{ C. W. Misner , K. S. Thorne, J.
A. Wheeler, \textit{``Gravitation''}, Feeman and Company (1973).}

{[}19{]} C. Corda, Eur. Phys. J. C 65 1-2, 257 (2010).

{[}20{]} S. Capozziello, C. Corda and M. F. De Laurentis, Phys. Lett.
B 669, 255 (2008). 

{[}21{]} C. Corda, Astropart. Phys. 30, 209 (2008).

{[}22{]} C. Corda and H. J. Mosquera Cuesta, Europhys. Lett. 86, 20004
(2009).

{[}23{]} S. Capozziello and C. Corda, Int. J. Mod. Phys. D 15, 1119
(2006). 

{[}24{]} S. Capozziello, C. Corda and M. F. De Laurentis, Mod. Phys.
Lett. A 22, 35, 2647 (2007).

{[}25{]} J. R. Smith, for the LIGO Scientific Collaboration, Class.
Quant. Grav. 26, 114013 (2009).

{[}26{]} R. A. Hulse and J.H. Taylor, Astrophys. J. Lett. 195, 151
(1975). 

{[}27{]} F. A. E. Pirani, Phys. Rev. 105, 1089 (1957).

{[}28{]} C. Corda, Astropart. Phys. 34, 412\emph{ }(2011).

{[}29{]}\foreignlanguage{italian}{ C. L. Bennett et al., }Astrophys.
J.\foreignlanguage{italian}{ Suppl. Ser. 148, 1 (2003).}

{[}30{]}\foreignlanguage{italian}{ D.N. Spergel et al., }Astrophys.
J.\foreignlanguage{italian}{ Suppl. Ser. 148, 195 (2003).}

{[}31{]} C. Corda and H. Mosquera Cuesta, Astropart. Phys. 34, 7,
587 (2011).

{[}32{]} C. Corda, Gen. Rel. Grav. 42, 1323 (2010).

{[}33{]} C. Corda, S. Capozziello, and M. F. De Laurentis, AIP Conf.
Proc. 966, 257 (2008).

{[}34{]} S. Capozziello, C. Corda and M. F. De Laurentis, Mod. Phys.
Lett. A 22, 15, 1097 (2007).

{[}35{]} C. Corda, Proceedings of the Workshop \textquotedbl{}\emph{Cosmology,
the Quantum Vacuum and Zeta Functions. A workshop with a celebration
of Emilio Elizalde's sixtieth birthday}\textquotedbl{}, S. D. Odintsov,
D. Sàez-Gòmez and S. Xambò-Descamps Editors, Springer Proceedings
137, 149 (2011).

{[}36{]} P. D. Scharre and C. M. Will, Phys. Rev. D 65, 042002 (2002).

{[}37{]} C. Corda, Phys. Rev. D 83, 062002 (2011).

\section*{Appendix: Gravitational waves in Extended Theories of Gravity. Quadrupole,
dipole and monopole modes. Potential detection}

The linearization process of the extended Einstein field equations
(\ref{eq: Einstein modified}), which has been developed in Refs.
{[}8, 20, 23, 28, 37{]}, enables GWs (again we choose the $z$ coordinate
like direction of propagation) 
\begin{equation}
h_{\mu\nu}(t,z)=h_{+}(t-z)e_{\mu\nu}^{(+)}+h_{\times}(t-z)e_{\mu\nu}^{(\times)}+h_{c}(t-v_{G}z)\eta_{\mu\nu}.\label{eq: perturbazione totale}
\end{equation}

The term $h_{+}(t-z)e_{\mu\nu}^{(+)}+h_{\times}(t-z)e_{\mu\nu}^{(\times)}$
describes the two standard polarizations of GWs which are present
in standard General Relativity too, while the term $h_{c}(t-v_{G}z)\eta_{\mu\nu}$
is the third polarization which is generated by perturbations of the
intrinsic curvature. The line element which arises from this third
polarization can be always put in a conformally flat form {[}3, 8,
20, 23, 28, 35{]}: 
\begin{equation}
ds^{2}=[1+h_{c}(t-v_{G}z)](-dt^{2}+dz^{2}+dx^{2}+dy^{2}).\label{eq: metrica puramente scalare}
\end{equation}

$v_{G}$ in Eq. (\ref{eq: metrica puramente scalare}) is the wave's
group velocity {[}3, 8, 20, 23, 28, 35{]}. In a particular case the
third mode can be massless and Eq. (\ref{eq: mass production}) becomes
Eq. (\ref{eq: zero-mass}) {[}3, 23, 28, 35{]}. In that case $v_{G}=1$.

In the framework of GWs, the most important difference between standard
General Relativity and Extended Theories of Gravity is the existence,
in the latter, of dipole and monopole radiation {[}36, 37{]}. In General
Relativity, for slowly moving systems, the leading multi-pole contribution
to gravitational radiation is the quadrupole one, with the result
that the dominant radiation-reaction effects are at order $(\frac{v}{c})^{5}$,
where $v$ is the orbital velocity. The rate, due to quadrupole radiation
in General Relativity, at which a binary system loses energy is given
by {[}36, 37{]}

\begin{equation}
(\frac{dE}{dt})_{quadrupole}=-\frac{8}{15}\eta^{2}\frac{m^{4}}{r^{4}}(12v^{2}-11\dot{r}^{2}).\label{eq:  Will}
\end{equation}

$\eta$ and $m$ are the reduced mass parameter and total mass, respectively,
given by $\eta=\frac{m_{1}m_{2}}{(m_{1}+m_{2})^{2}}$, and $m=m_{1}+m_{2}$.

$r,$ $v,$ and $\dot{r}$ represent the orbital separation, relative
orbital velocity and radial velocity respectively.

In of Extended Theories of Gravity, Eq. (\ref{eq:  Will}) is modified
by high-order corrections {[}36, 37{]}. Monopole radiation is also
predicted but in binary systems it contributes only to the high-order
corrections {[}36, 37{]}. The important modification in of Extended
Theories of Gravity is the additional energy loss caused by dipole
modes which are generated by perturbations of the intrinsic curvature
{[}8, 20, 23, 28, 37{]}. By analogy with electrodynamics, dipole radiation
is a $(v/c)^{3}$ effect, potentially much stronger than quadrupole
radiation. However, in Extended Theories of Gravity, the gravitational
\textquotedblleft{}\emph{dipole moment}\textquotedblright{} is governed
by the difference $s_{1}-s_{2}$ between the bodies, where $s_{i}$
is a measure of the self-gravitational binding energy per unit rest
mass of each body {[}36, 37{]}. $s_{i}$ represents the \textquotedblleft{}\emph{sensitivity}\textquotedblright{}
of the total mass of the body to variations in the background value
of the Newton constant, which is a function of the intrinsic curvature
(expressed in terms of scalar fields or of high-order terms in the
curvature invariants) {[}8, 20, 23, 28, 37{]}: 

\begin{equation}
s_{i}=\left(\frac{\partial(\ln m_{i})}{\partial(\ln G)}\right)_{N}.\label{eq: Will 2}
\end{equation}

\emph{$G$} is the effective Newtonian constant at the star, which
depends on the particular Extended Theory of Gravity, and the subscript
$N$ denotes holding baryon number fixed. 

Defining $S\equiv s_{1}-s_{_{2}}$, to first order in $\frac{1}{\omega}$
the energy loss caused by dipole radiation in Scalar-Tensor Gravity
is given by {[}36, 37{]} 
\begin{equation}
(\frac{dE}{dt})_{dipole}=-\frac{2}{3}\eta^{2}\frac{m^{4}}{r^{4}}(\frac{S^{2}}{\omega}),\label{eq:  Will 3}
\end{equation}

where $\omega$ is the famous Brans-Dicke parameter {[}36, 37{]}. 

In Extended Theories of Gravity, the sensitivity of a black hole is
$s_{BH}=0.5$ {[}36, 37{]}, while the sensitivity of a neutron star
varies with the equation of state and mass. For example, $s_{NS}\approx0.12$
for a neutron star of mass order $1.4M_{\circledcirc}$, being $M_{\circledcirc}$
the solar mass {[}36, 37{]}. 

Binary black-hole systems are not at all promising for studying dipole
modes because $s_{BH1}-s_{BH2}=0,$ a consequence of the no-hair theorems
for black holes {[}36, 37{]}. In fact, black holes radiate away any
scalar field, so that a binary black hole system in Extended Theories
of Gravity behaves as if General Relativity. Similarly, binary neutron
star systems are also not effective testing grounds for dipole radiation
{[}36, 37{]}. This is because neutron star masses tend to cluster
around the Chandrasekhar limit of $1.4M_{\circledcirc}$, and the
sensitivity of neutron stars is not a strong function of mass for
a given equation of state. Thus, in systems like the binary pulsar,
dipole radiation is naturally suppressed by symmetry {[}36, 37{]}.
Hence the most promising systems are mixed: black hole - neutron star,
black hole - white dwarf, or neutron star - white dwarf. 

The emission of monopole gravitational radiation in Extended Theories
of Gravity is very important in the collapse of quasi-spherical astrophysical
objects because in this case the energy emitted by quadrupole modes
can be neglected {[}28, 37{]}. In the formation of a neutron star,
monopole waves interact with the detectors as well as quadrupole ones
{[}28, 37{]}. 

Resuming, there are two potential sources of potential detectable
third mode: mixed binary systems of very compact stars and gravitational
collapse of quasi-spherical astrophysical objects.

The second source looks propitious because in such a case the energy
emitted by quadrupole modes can be neglected {[}28, 37{]} (in the
sense that the monopole modes largely exceed the quadrupole ones.
In fact, if the collapse is completely spherical, the quadrupole modes
are totally removed {[}18, 28, 37{]}). In that case, only the motion
of the test masses due to the third mode $h_{c}$ has to be analysed.
An interesting case is the formation of a neutron star through a gravitational
collapse {[}28, 37{]}. A collapse occurring closer than 10 kpc from
us (half of our Galaxy) needs a sensitivity of  $3\times10^{-23}\mbox{ }\sqrt{Hz}$
at $800\mbox{ }Hz$ (which is the characteristic frequency of such
events) to potential detect the strain which is generated by the $h_{c}$
mode in the arms of LIGO {[}28, 37{]}. At the present time, the sensitivity
of LIGO at about $800\mbox{ }Hz$ is $10^{-22}\mbox{ }\sqrt{Hz}$
while the sensitivity of the Enhanced LIGO Goal is predicted to be
$8\times10^{-22}\mbox{ }\sqrt{Hz}$ at $800\mbox{ }Hz$ {[}25, 28,
37{]}. Then, for a potential detection of the third mode $h_{c}$,
we have to hope in Advanced LIGO Baseline High Frequency and/or in
Advanced LIGO Baseline Broadband. In fact, the sensitivity of these
two advanced configuration is predicted to be $6\times10^{-23}\mbox{ }\sqrt{Hz}$
at $800\mbox{ }Hz$ {[}25, 28, 37{]}. 

The sensitivities of interferometric GWs detectors are usually expressed
in $strain\times\mbox{ }\sqrt{Hz}$ {[}25{]}. The \emph{strain} is
a non-dimensional quantity which represents the ratio between the
amplitude of the oscillation of the test masses (the mirror and the
beam-splitter of the interferometer) and the distance between them.
For astrophysical and cosmological sources the strain is always $\leq10^{-21}$
even for GWs in standard General Relativity. This implies that interferometers
having a distance between test masses of order kilometres should measure
amplitudes of oscillation shorter than $10^{-18}$ meters {[}25{]},
and this makes the potential GW detection extremely difficult to be
realized. In fact, a direct GW detection has not yet achieved. We
hope in an improved sensitivity of advanced projects.
\end{document}